\documentstyle[11pt,twoside,epsf,newpasp]{article}
\newcommand{\flux}{\rm erg\ cm^{-2}\ s^{-1}}

\begin{document}
\title{X-ray afterglows and spectroscopy of Gamma-Ray Bursts}
 \author{Luigi Piro}
\affil{Istituto Astrofisica Spaziale e Fisica Cosmica, CNR, Via
Fosso Cavaliere 100, Rome, Italy}

\begin{abstract}
I will review the constraints set by X-ray measurements of
afterglows on several issues of GRB, with particular regard to the
fireball model, the environment, the progenitor and  dark GRB.
\end{abstract}

\section{Introduction}

This conference took place few months after the switch-off of
BeppoSAX (\cite{psbsax95}), on April 30, 2002. Launched on April
30, 1996, this mission carried out  observations of all classes of
X-ray sources during its  operative lifetime of 6 years. A total
of 62 Msec of pointed observations with its Narrow Field
Instruments (NFI) were carried out. A substantial fraction (about
50\%) of the total observing programme was devoted to observations
of compact galactic sources and AGN, i.e. the classes of sources
mostly suited to the exploitation of the broad band spectral
coverage of BeppoSAX NFI (0.1-200 keV).

The other strong asset  of the mission was the capability of
discover and carry out deep observations of transient phenomena in
the sky. This was assured by wide field X-ray and gamma-ray
monitors (Wide Field Cameras, WFC, and Gamma-Ray Burst Monitor,
GRBM) coupled with a high level of flexibility of  ground
 scientific operations in carrying out fast Target of
Opportunity Observations (TOO) with NFI. In fact, a substantial
part of the program was devoted to such observations: about 190
NFI observations (corresponding to a total of 7.2 Msec), out of
which 2.2 Msec  on Gamma-Ray Bursts. Turning then to GRB, 56 GRB
(including 8 X-ray rich GRB) were localized by wide field
instruments and their position distributed within few hours. 38
GRB were observed with fast TOO observations (from 5 hrs to 1 day)
with NFI. The first GRB observation took place on July 20, 1996,
during the scientific verification phase (Piro et al, 1997), and
the last one just the last days of operations. The most famous
events were GRB970228, that led to the discovery of the first
X-ray and optical afterglows (\cite{cfh+97,vgg+97}),  GRB970508
 whose precise and fast localization (\cite{paa+98}) allowed the
first determination of  distance and the discovery of the first
radio afterglow and fireball observational evidence
(\cite{mdk+97,fkn+97}), and GBR980425 (\cite{paa+00}), with its
association with SN1998bw (\cite{gvv+98}).

In recent years, most of the research activities in the field have
focussed on 3 main topics.

\begin{itemize}

\item Progenitor and central engine.

\item Origin of dark GRB, X-ray flashes and short GRB

\item Cosmology with GRB

\end{itemize}

Those areas of research are closely intertwined. The origin of
dark GRB or X-ray flashes could have relevant implications both on
the progenitor/central engine and on cosmological studies. The
nature of the progenitor is then relevant to cosmological studies
with GRB, because of their possible use as  tracers of
star-formation in the Universe.  In this review I will focus on
the impact of X-ray measurements on the fireball model, the
environment and the origin of the progenitor,  and on dark GRB.

\begin{figure}
\plottwo{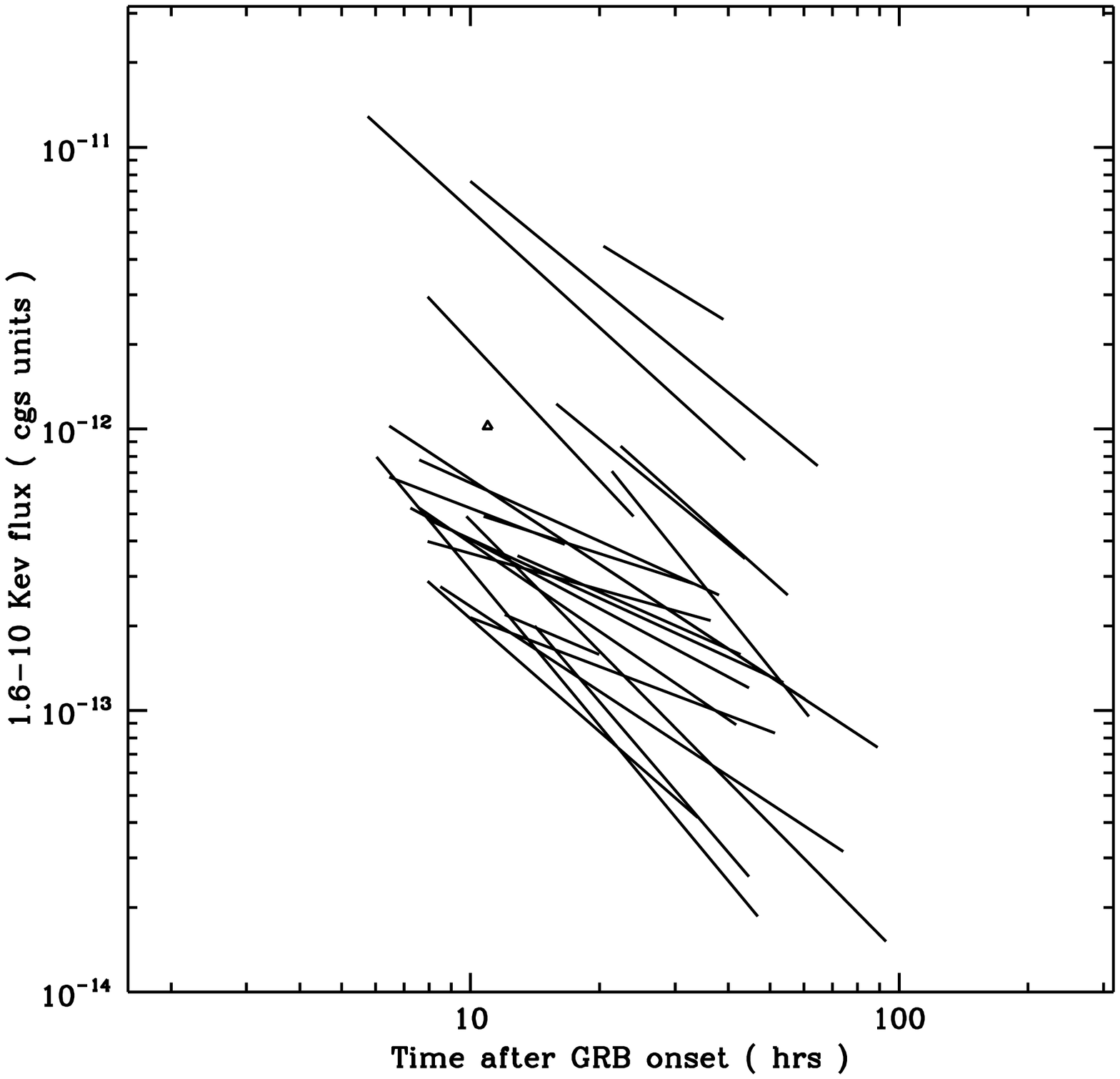}{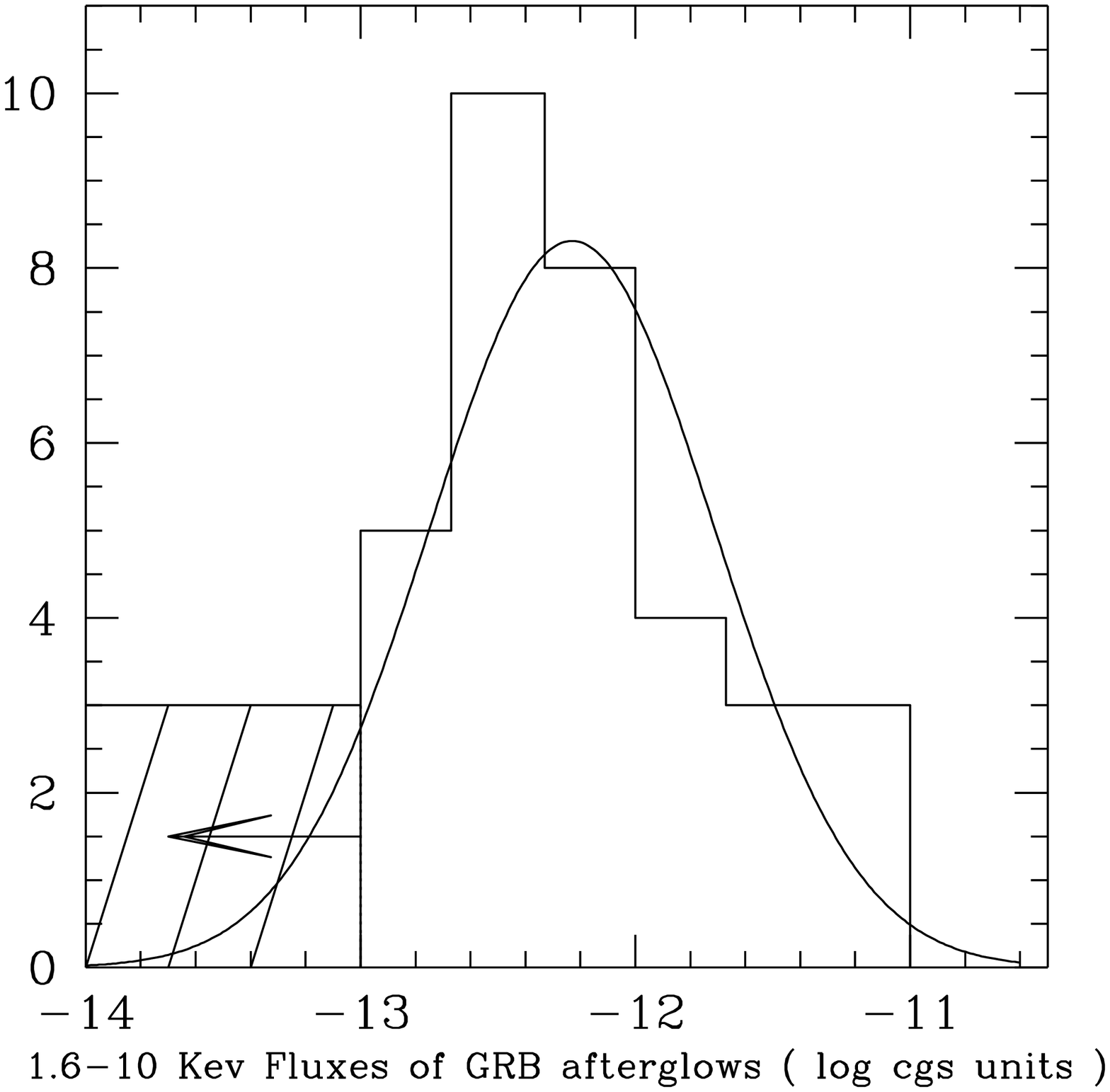}
\caption{Light curves (best fit power laws: left panel) and
distribution of F(1.6-10 keV) at 11 hrs (right panel) with
BeppoSAX}
\end{figure}

\section{The catalogue of BeppoSAX afterglow observations}
BeppoSAX has performed 39 follow-up observations of GRB, 38
following localizations by BeppoSAX wide field instruments, and
one (GRB000926) from an external trigger. We have considered here
36 observations, excluding the cases of GRB960720 (the first
 localization of a GRB by BeppoSAX when the TOO was performed one month after the burst),
GRB990705 (due to a contamination from off-axis strong source),
and GRB980425 (=Sn1998bw). Results from a subset on this sample
(31 GRB) have been published in De Pasquale et
al.(2003)\nocite{dpp+03}. The observations  started typically 8
hours after the burst (ranging from 5 hours to 1 day), usually
with a second observation taken 1-2 days after the burst. The
sample is constituted primarily by events triggered by the
gamma-ray burst monitor, but includes also all the X-ray rich GRB
and X-ray flashes triggered (in real time) by the X-ray Wide Field
Cameras. All  events were long GRB, with the shortest  lasting
about 2 seconds.

 The first result of the analysis is that
  the X-ray afterglow is a common feature in GRB.
  Only in  three cases we do not find any source in the WFC error box
(with an upper limit around $10^{-13} \flux$), while in three
other events a source is detected but with no significant fading
behaviour. Therefore X-ray afterglows are present in $\ga 83-92\%$
of the GRB. We will come back to this result in the context of
dark GRB in a following section.

\begin{figure}
\plottwo{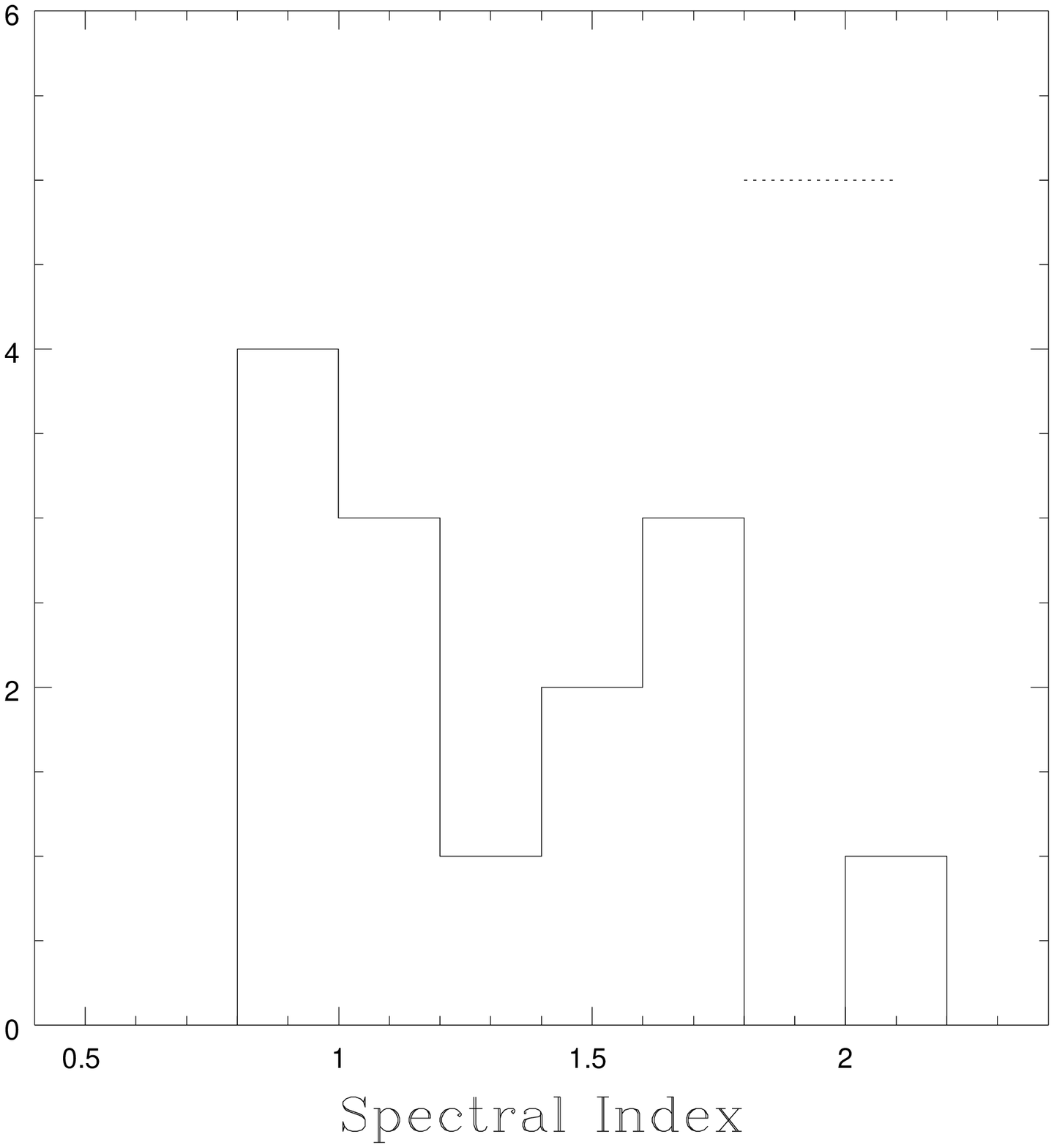}{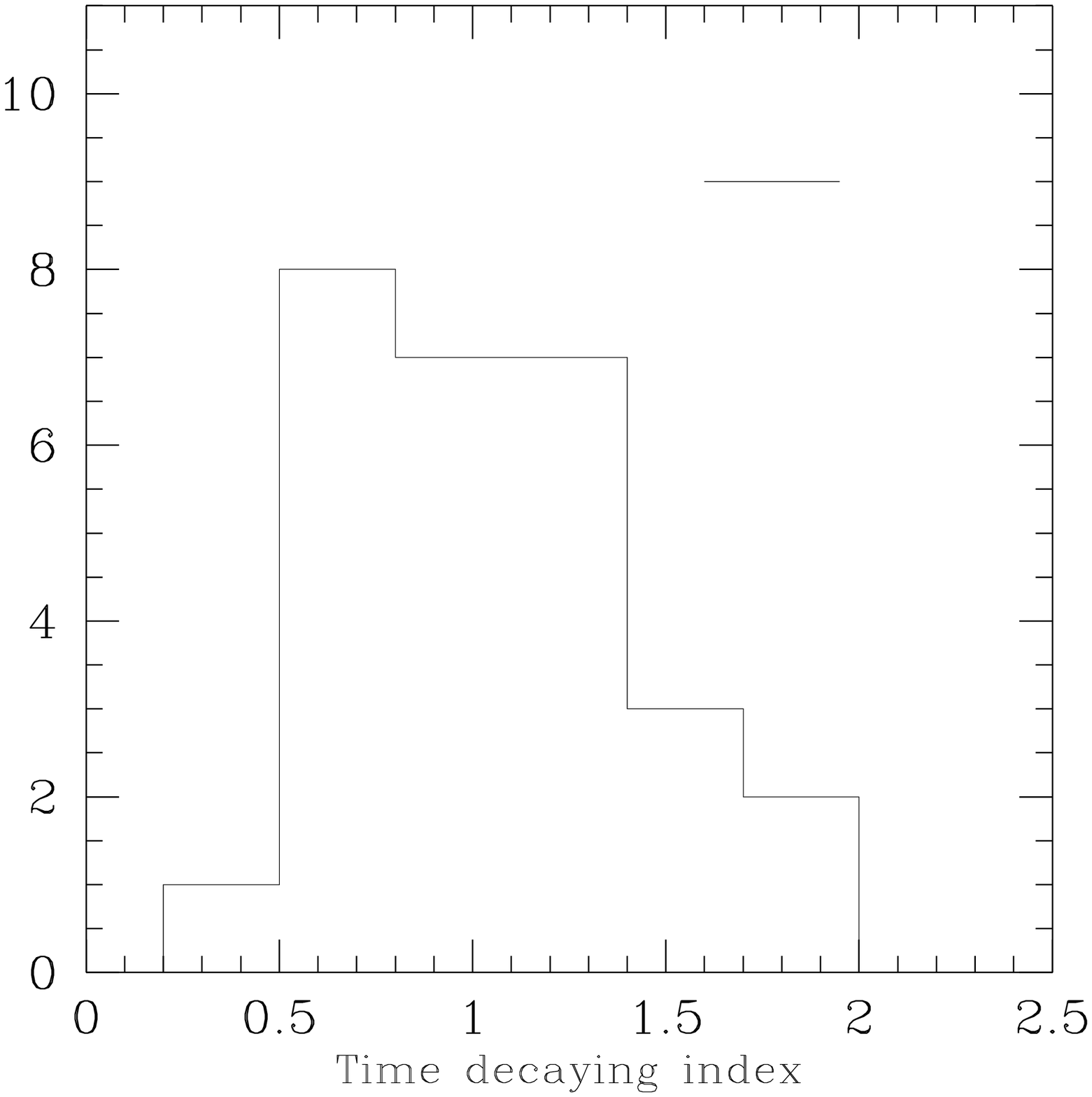}
\caption{Distribution of spectral and temporal indeces of
afterglows observed with BeppoSAX}
\end{figure}

\section{Implications on the fireball model and jet scenario}

We have characterized the temporal and spectral behaviour of the
afterglows with a power law model, including the absorption by our
Galaxy and at the source, as follows: $F(t,E)= F_0 exp({-\sigma
N_H})t^{-\delta}E^{-\alpha}$, where $\alpha$ is the energy
spectral index. In Fig.1 (left panel) we show a collection of
power law decay for afterglows in our sample.

We find, confirming the result reported in Piran et al. (2001)
\nocite{pkpp01}, -- that was based on a more limited sample of
events --, that the distribution of afterglow fluxes observed at
11 hours is very narrow (Fig.1, right panel). As argued by Piran
et al., this result implies that the kinetic energy of the
fireball in all GRB of this sample is also very narrowly
distributed, independently supporting the result by Frail et al.
(2001) \nocite{fks+01} of a universal energy reservoir on GRB. We
note, however, that the analysis of Piran et al provides only the
width of the distribution, and not the absolute value of the
energy.

Now we focus on the distribution of temporal and spectral indeces
and the implications on the fireball model. We find that both
indeces distribution are narrow, clustering around the average
values $\alpha=1.13\pm0.07$ and $\delta=1.2\pm0.1$ (Fig.2). The
 determination of these values allows a comparison with
the expectation from different realization of the fireball model,
in which the two indeces are related each other through the so
called closure relations (\cite{sph99,cl99}). We consider fireball
expansion in a constant density medium (ISM) or in a wind-like
medium, with a density profile following $r^{-2}$ (wind) both for
a spherical and a collimated (jet) flow. The results are
summarized in table.1, where the closure parameter $C$ (that has
to be consistent with 0) is given for these different cases. The
following conclusion are derived for the average properties of the
fireball in the time frame from few hours to 1-2 days.

\begin{itemize}

\item The X-ray emission is generated from electrons in the
cooling regime (i.e. the cooling frequency $\nu_c$ is below the
X-ray range)

\item The index of the electron distribution is $p=2.26\pm0.14$

 \item The fireball expansion for t$\la$2 days is consistent with
 spherical outflow, either in ISM or wind.  This allows us to set
 a lower limit to the collimation angle of the jet:
 $\theta>11\deg (n/E_{iso,52})^{1/8} (t/2days)^{3/8}$
 (vs the average of 6$\deg$ from Frail et al. 2001)

 \end{itemize}

\begin{table}
\caption{Constraints on the fireball model from X-ray afterglows}
\begin{tabular}{llll}
\tableline
Regime & ISM & Jet & Wind \\
\hline
$\nu<\nu_c$ & $C=\delta-\frac{3}{2}\alpha$ & $C=\delta-2\alpha-1$ & $C=\delta-\frac{3}{2}\alpha-\frac{1}{2}$  \\
$\alpha=\frac{p-1}{2}$ & $C=-0.49\pm0.14$ & $C=-2.03\pm0.17$ &
$C=-1.0\pm0.14$  \\[1pt]
\hline
$\nu>\nu_c$ & $C=\delta-\frac{3}{2}\alpha+\frac{1}{2}$ & $C=\delta-2\alpha$ & $C=\delta-\frac{3}{2}\alpha+\frac{1}{2}$  \\
$\alpha=\frac{p}{2}$ & C=$0.0\pm0.14$ & $C=-1.0\pm0.17$ & $C=0.0\pm0.14$  \\[1pt]
 \hline \hline
\end{tabular}
\end{table}

\section{Broad and narrow X-ray features: environment and progenitors}
The GRB and its afterglow are very well explained by the fireball
model, in which a highly relativistic outflow from the central
source produces the observed emission. On the other hand, this
process essentially loses "memory" of the central source: the
shocks that are thought to produce  GRB and afterglow photons take
place over a distance scale that is about 10 orders of magnitude
greater than the size of the central source. In addition, this is
almost independent of the details of the central source, depending
primarily on basic parameters as the total energy, the collimation
angle of the outflow (jet), the fraction of energy in relativistic
electrons and magnetic fields and the density of the external
medium.

A very effective method to gather information about the progenitor
is to study line features produced in the environment of the GRB.
The iron line is an ubiquitous feature  in all families of X-ray
sources (\cite{piro93}). Most of the searches of features in GRB
have been therefore concentrated at the energies of this element,
i.e. (at rest frame) 6.4 to 6.9 keV for $K_\alpha$ lines from
neutral to H-like ions, 9.3 keV for the recombination edge in
emission from H-like ions and 7.1 keV, the energy of the
absorption edge from neutral iron in absorption. So far there are
6 independent measurements of iron features from 4 different
satellites (see Piro (2003) \nocite{piro03} and references therein
plus the recent case of GRB010220 (\cite{wro+02})). While each
single measurement is not of overwhelming statistical
significance, the overall scenario is rather compelling. There are
so far four burst with an independent redshift measurement from
optical spectra.  In three of these events the emission features
detected in the afterglow phase
 are consistent with highly ionized
iron, while in one case there is evidence of a transient
absorption edge during the main GRB pulse (\cite{afv+00}).

In the {\it distant reprocessor scenario} the line-emitting medium
is external to the fireball region, as suggested by the presence
of the absorption edge. In the early phase of the burst this
medium is still to be completely ionized by the GRB photons, thus
producing an absorption edge from neutral iron. As the ionization
front reaches out the external border of the medium, this becomes
completely ionized (\cite{pl98,bdcl99}), thus explaining the
disappearence of the absorption edge. On a time scale given by the
recombination time, electrons start to recombine on ionized iron,
thus producing the emission line and recombination edge observed
in the afterglow phase. In an alternative model
(\cite{mr_line01}), it is assumed that the central source, after
the event producing the GRB, continues its activity - at lower
power -, heating and ionizing a close-by  line emitting medium
({\it local reprocessor scenario}). The progenitor is likely a
massive star that undergoes a core-collapse supernova explosion
(collapsar: \cite{w01}). In the distant reprocessor scenario, this
explosion takes place about a month before the event leading to
the GRB (\cite{vs_supra99}), and are the Supernova ejecta
illuminated by
 X-ray photons of the gamma-ray burst that produce the lines. This is also
 consistent with the line width observed in GRB991216 (\cite{pgg+00}), that corresponds
 to an outflow velocity of 10\% of the speed of light,
 as typically observed in Supernovae.
 In the local reprocessor scenario the two events are almost
simultaneous. In addition to Fe features, recent detections of
soft X-ray lines by ionized elements as S, Si, Mg (\cite{r+02})
supports the association of GRB with SN-like explosions. In
particular, those lines are blue-shifted with respect to the
rest-frame energies by about 10\% of the speed of light.

A point to be noted regards the origin of iron in the two
scenarios. In the case of distant reprocessor, a large mass of
iron is required, $\approx 0.05 M_{\odot}$. This is consistent
with the mass of iron group elements  ejected in SN explosion.
Because iron is actually the end result of the decay chain
Co56(6.1 days)-Ni56(78.8 days)-Fe56 decay, a minimum delay between
the SN explosion and the GRB of 2 months is required. In the case
of the local reprocessor scenario, a modest amount of iron
($<10^{-8} M_{\odot}$) suffices. In the framework of the collapsar
model  this material is advected from the iron core when the jet
and its associated cocoon propagate around the rotation axis of
the star and break out at the surface of the star. The energy in
the cocoon ($\approx 10^{51} erg$) could be released in few hours
after the burst and be the source of ionization and heating of the
line-emitting material.

\section{X-ray absorbers and environment}
The spectra of X-ray afterglows show, in several cases, an
intrinsic absorption in the range $10^{21-22} cm^{-2}$
(\cite{dpp+03,s+03}). This is consistent with the column density
measured in star-forming Giant Molecular Clouds in our Galaxy
(\cite{srb+87}),  strengthening the connection of GRB with star
formation sites.

Absorption data in the prompt phase are still very sparse. There
are a few GRB with an absorption column density $N_H\ga
10^{23}cm^{-2}$ (e.g. \cite{fac+00,iz+01,gfm+03}). On the
contrary, the column density measured in the afterglow phase {\it
in the same burst} is one order of magnitude less. This behaviour
is expected if the absorbing medium with a density typical of a
GMC ($n\approx10^2-10^5 cm^{-3}$) is indeed lying within a few
parsec from the GRB. In such a case it will be ionized by the GRB
photons, becoming effectively transparent in the afterglow phase
(\cite{lpg02}). Piro et al. (2002) \nocite{pfg+02} have noticed
that the absorber spectrum in the afterglow of GRB000210 is
consistent with a medium in a low ionization stage, suggesting the
possibility that the absorber is actually condensed in
high-density ($n\approx 10^9 cm^{-3}$) clouds. The large  column
density in the prompt phase could then be explained by the very
small fireball region visible to the observer  being fully covered
by a single cloud. As the visible region of the fireball
increases, it will become {\it partially covered} by the cloud
ensemble, producing again a reduction in the effective column
density. This scenario can also account for the erratic variations
of the column density reported during the prompt phase of
GRB010222 (\cite{iz+01}). Clearly, high quality spectra are needed
to progress beyond the simple uniform absorber model that is
fitting the present data, allowing detailed test of partially
covered and photoionized absorbers.

\section{The X-ray view of dark GRB}
One of the most intriguing issues on recent research of GRB
regards the origin of the so-called dark GRB. We have mentioned
above that about 90\% of the GRB do show an X-ray afterglow. On
the contrary,  only about 40\% of them  have an optical afterglow
(\cite{dpp+03}).  There has been considerable discussions on
whether this effect is due to an observational bias or not. The
first point to be stressed is that the optical upper limits on
these events lie on average two magnitudes below the average
magnitudes measured for events with optical transients
(OTGRB)(\cite{lcg02}). However, the detection of a few optical
transients with magnitudes below some of the upper limits on dark
GRB (e.g. \cite{bkb+02})  suggests that some of the dark GRB are
indeed a faint end extension of OTGRB, rather than a separate
class of events.

Apart from semantic consideration, the origin of this behaviour
would still be not-trivial. For example, it could be due to an
intrinsic property (underluminous events), a distance effect (but
at z$\la$5, since for higher redshift the optical should be almost
completely absorbed, see below) or a fast decay, as that expected
for a highly collimated jet. {\it In all these cases the afterglow
flux should scale of the same factor at all wavelengths}.

{\it On the contrary, the optical flux of a GRB at z$\ga$5 or of a
GRB in a dusty star forming region should be depleted not only in
absolute magnitude  but also with respect to other wavelengths}.
We have therefore carried out a study of dark GRB vs OTGRB
comparing their X-ray vs optical fluxes (see also De Pasquale et
al, this conference). The results can be summarized as follows.

\begin{itemize}
\item The X-ray flux of  afterglows of dark GRB's is on average a
factor of 6 lower than that of OTGRB.

\item In 75\% of dark GRB's, the upper limits on the
optical-to-X-ray flux ratio ($f_{OX}$)  are consistent with the
ratio observed in OTGRB. This population of events is therefore
consistent with being   OTGRB going undetected in the optical
because searches were not fast or deep enough.

\item However, for about 25\% of dark GRB, $f_{OX}$ is at least a
factor 5-10 lower than the average value observed in OTGRB, and
also lower than the smallest observed $f_{OX}$. Furthermore, the
optical upper limits on these events are also lower than the
faintest optical afterglow. These GRB cannot be therefore
explained as dim OTGRB's, and we refer to them as {\it truly dark
or optically depleted GRB}.
\end{itemize}

We stress that the upper limit on $f_{OX}$ for optically depleted
GRB is model-independent, being derived by a comparison with the
optically bright GRB, where the $f_{OX}$ distribution is rather
narrow,  clustering around the average value within a factor of 2
(the 1 sigma width). It is then worth mentioning that a similar
value on this  limit  has been derived in two dark GRB
(\cite{dfk+01,pfg+02}) by modelling the broad band data via the
standard fireball model. Both these events have been associated
with host galaxies at z$\la$5, leading to the conclusion that the
optical is depleted by dust in star-forming region. Indeed, one of
these two objects (GRB000210) is also included in our sample.

 This association does not exclude that other optically depleted
 GRB are indeed at z$\ga$5. Actually, Bromm \& Loeb (2002)
\nocite{bl02}have estimated that more than
 20-30\% of GRB should lie at z$\ga$5. Indeed, we find that
 the average X-ray afterglow flux of optically
 depleted GRB's is 5 times lower than OTGRB's, an effect that can
 be straightforwardly attributed to distance.
 We point out, however, that this effect could also be explained in the
 obscuration scenario, assuming that dark GRB are less collimated than OTGRB's
 while retaining a similar total energy (\cite{ry01}).

Since most of the redshift of GRB are derived from optical
spectra, there is a strong observational bias against high-z GRB.
This limitation can be overcome only by  X-ray spectroscopy (X-ray
redshift) or far infrared measurements.


\begin{thebibliography}{}

\bibitem[{Amati} {\it et al.}  2000]{afv+00}
{Amati}, L. {\it et al.}  2000, Science, 290, 953.

\bibitem[{Berger} {\it et al.}  2002]{bkb+02}
{Berger}, E., {Kulkarni}, S.~R., {Bloom}, J.~S., {\it et al.}  2002, ApJ, 581,
  981.

\bibitem[{Boettcher} {\it et al.}  1999]{bdcl99}
{Boettcher}, M., {Dermer}, C.~D., {Crider}, A.~W., and {Liang}, E. .~P. 1999,
  A\&A, 343, 111.

\bibitem[Bromm \& Loeb 2002]{bl02}
Bromm, V. and Loeb, A. 2002, ApJ, 575, 111.

\bibitem[{Chevalier} \& {Li} 1999]{cl99}
{Chevalier}, R.~A. and {Li}, Z.-Y. 1999, ApJ, 520, L29.

\bibitem[{Costa} {\it et al.}  1997]{cfh+97}
{Costa}, E. {\it et al.}  1997, Nature, 387, 783.

\bibitem[{De Pasquale} {\it et al.}  2003]{dpp+03}
{De Pasquale}, M., Piro, L., Perna, R., {\it et al.}  2003, ApJ, 5092, 1018.

\bibitem[Djorgovski {\it et al.}  2001]{dfk+01}
Djorgovski, S.~G., , Frail, D.~A., Kulkarni, S.~R., Bloom, J.~S., Odewahn,
  S.~C., and Dierks, A. 2001, ApJ, 562, 654.

\bibitem[{Frail} {\it et al.}  1997]{fkn+97}
{Frail}, D.~A., {Kulkarni}, S.~R., {Nicastro}, S.~R., {Feroci}, M., and
  {Taylor}, G.~B. 1997, Nature, 389, 261.

\bibitem[Frail {\it et al.}  2001]{fks+01}
Frail, D.~A. {\it et al.}  2001, ApJ, 562, L55.

\bibitem[{Frontera} {\it et al.}  2000]{fac+00}
{Frontera}, F. {\it et al.}  2000, ApJS, 127, 59.

\bibitem[{Galama} {\it et al.}  1998]{gvv+98}
{Galama}, T.~J. {\it et al.}  1998, Nature, 395, 670.

\bibitem[Guidorzi {\it et al.}  2003]{gfm+03}
Guidorzi, C. {\it et al.}  2003, A\&A, 401, 491.

\bibitem[{in' t Zand} {\it et al.}  2001]{iz+01}
{in' t Zand}, J. . J.~M. {\it et al.}  2001, ApJ, 559, 710.

\bibitem[Lazzati, Covino \& Ghisellini 2002]{lcg02}
Lazzati, D., Covino, S., and Ghisellini, G. 2002, MNRAS, 583.

\bibitem[Lazzati, Perna \& Ghisellini 2002]{lpg02}
Lazzati, D., Perna, R., and Ghisellini, G. 2002, MNRAS, 325, L19.

\bibitem[{M{\'e}sz{\' a}ros} \& {Rees} 2001]{mr_line01}
{M{\'e}sz{\' a}ros}, P. and {Rees}, M.~J. 2001, ApJ, 556, L37.

\bibitem[Metzger {\it et al.}  1997]{mdk+97}
Metzger, M.~R., Djorgovski, S.~G., Kulkarni, S.~R., Steidel, C.~C., Adelberger,
  K.~L., Frail, D.~A., Costa, E., and Fronterra, F. 1997, Nature, 387, 879.

\bibitem[{Perna} \& {Loeb} 1998]{pl98}
{Perna}, R. and {Loeb}, A. 1998, ApJ, 501, 467.

\bibitem[Pian {\it et al.}  2000]{paa+00}
Pian, E., Amati, L., Antonelli, L.~A., {\it et al.}  2000, ApJ, 536, 778.

\bibitem[{Piran} {\it et al.}  2001]{pkpp01}
{Piran}, T., {Kumar}, P., {Panaitescu}, A., and {Piro}, L. 2001, ApJ, 560,
  L167.

\bibitem[Piro 1993]{piro93}
Piro, L. 1993, in { UV and X-ray spectroscopy of Laboratory and Astrophysical
  Plasmas}, ed.\ S.~Kahn E.~Silver, Cambridge Uni. Press, 448.

\bibitem[Piro 2003]{piro03}
Piro, L. 2003, in { Proceedings of GRB and afterglow astronomy 2001}, 372.

\bibitem[Piro {\it et al.}  1998]{paa+98}
Piro, L., Amati, L., Antonelli, L.~A., {\it et al.}  1998, A\&A, 331, L41.

\bibitem[Piro {\it et al.}  2002]{pfg+02}
Piro, L., Frail, D., Gorosabel, J., {\it et al.}  2002, ApJ, 577, 680.

\bibitem[{Piro} {\it et al.}  2000]{pgg+00}
{Piro}, L. {\it et al.}  2000, Science, 290, 955.

\bibitem[Piro, Scarsi \& Butler 1995]{psbsax95}
Piro, L., Scarsi, L., and Butler, R. 1995, SPIE, 169.

\bibitem[{Reeves} {\it et al.}  2002]{r+02}
{Reeves}, J.~N. {\it et al.}  2002, Nature, 415, 512.

\bibitem[Reichart \& Yost 2001]{ry01}
Reichart, D.~E. and Yost, S.~A. 2001, ApJ, submitted; astro-ph/0107545.

\bibitem[{Sari}, {Piran} \& {Halpern} 1999]{sph99}
{Sari}, R., {Piran}, T., and {Halpern}, J.~P. 1999, ApJ, 519, L17.

\bibitem[Solomon {\it et al.}  1987]{srb+87}
Solomon, P.~M., Rivolo, A.~R., Barrett, J., and Yahil, A. 1987, ApJ, 319, 730.

\bibitem[Stratta {\it et al.}  2003]{s+03}
Stratta, G. {\it et al.}  2003, ApJ.
\newblock submitted.

\bibitem[{van Paradijs} {\it et al.}  1997]{vgg+97}
{van Paradijs}, J. {\it et al.}  1997, Nature, 386, 686.

\bibitem[{Vietri} \& {Stella} 1999]{vs_supra99}
{Vietri}, M. and {Stella}, L. 1999, ApJ, 527, L43.

\bibitem[Watson {\it et al.}  2002]{wro+02}
Watson, D., Reeves, J.~N., Osborne, J., O'Brien, P.~T., Pounds, K.~A., Tedds,
  J.~A., Santos-Lle, M., and Ehle, M. 2002, A\&A, 393, L1.

\bibitem[Woosley 2001]{w01}
Woosley, S.~E. 2001, in { GRBs in the Afterglow Era}, ed.\ E. Costa, F.
  Frontera, and J. Hjorth, ESO-Springer, 258.

\end{thebibliography}

\end{document}